\documentclass{article}

\usepackage{amsmath}

\title{p-Norm Flow Optimization in a Network}
\author{Sina Z. Anaraki, Mehdi Kalantari}
\date{}
\begin{document}
\maketitle
In this paper we study information flow paths in a data network, where traffic generated by servers ( or sources) takes a multi-hop path in order to reach its clients (destinations).  Each node in the middle of this multi-hop path should route the incoming traffic and the traffic generated by itself to the next hop in such a way that the traffic reaches its destination while avoiding congestion in the links. For simplicity, we will assume the network only carries single commodity traffic, i.e., all of the traffic should be routed to a single destination.

Let us assume that the data network is modeled by a connected graph $G(V,E)$ where the set of vertices, $V$ ($|V| = N$) represents the nodes (or data servers) and the set of edges, $E$ ($|E| = M$) represents the links. For our analysis, we assume that the destination for all of the traffic in the network is node $N$.  Two nodes are called adjacent if they are connected by an edge.
A \emph{cut-set} is a set of edges that if they are removed from the graph it becomes disconnected. A \emph{minimum cut-set} is a cut-set with minimum cardinality. 
The throughput of a network with arbitrary sources and destinations is limited by the capacity of the links in the minimum cut-set of the network. If traffic is not routed carefully, when the network is under heavy traffic, some of these links may become congested while other are under-utilized. 
It is desirable to route the traffic such that each link in the minimum cut-set achieves its maximum capacity. When the capacity of all of these links are the same, an optimal routing would route the traffic on the bottleneck links such that the traffic is distributed evenly among them. 

We arbitrarily assign a direction to each edge $e_m$ such that if it connects nodes $r$ and $s$ where $r<s$, then the the direction of $e_m$ is from $r$ to $s$. Let us denote the traffic load on edge $e_m$ with $I_m$.  Note that if traffic flows from node $r$ to node $s$ then $I_m$ is positive and it is negative if traffic flows in opposite direction. For each node $n$, the sum of incoming traffic load (from adjacent nodes with lower index ) and the traffic generated by the node itself ($T_n$) is equal to the sum of outgoing traffic (to adjacent nodes with higher index). These conservation of traffic constraints are written for nodes $1\ldots N-1$ in matrix form as
\begin{align}
C \cdot I = T
\end{align}
where $C$ is a $(N-1)\times M$ matrix with elements in $\{-1,0,+1\}$, I is a $M\times 1$ vector of traffic loads $I_m$ and $T$ is a $N-1\times 1$ vector of traffic sources $T_n$. For $[C]_{ij}$ we have
\begin{align}
[C]_{i,j} =\left\{
		\begin{array}{lll}
			-1 & \mbox{if } e_j \mbox{ connects node } i \mbox{ to a node with lower index} \\
			0 & 1 \mbox{if } e_j \mbox{ does not start or end with node } i \\
			+1 &   \mbox{if } e_j \mbox{ connects node } i \mbox{ to a node with higher index}
		\end{array}
	\right.
\end{align}

We define a cost function 
\begin{align}
J_p(I) =\| I\|_{p}^{p}= \sum_{m=1}^{M} I_m^{p}
\end{align}
and write the following optimization problem:

\begin{equation}\label{Eq:Opt}
	\begin{aligned}
		& \underset{I}{\text{minimize}} & & J_p(I) \\
		& \text{subject to} & & C\cdot I = T
	\end{aligned}
\end{equation}

%\textbf{Special Case} $p=2$:

In order to solve this problem in a distributed fashion we employ Sequential Quadratic Programming (SQP) in conjunction Jacobi iterations.

Let $i$ be a small perturbation in $I$ such that $C\cdot (I+i) = T$. Thus $C\cdot i = 0$ . We also have
\begin{align}
J_p(I+i) &= \sum_{m=1}^{M} (I_m+i_m)^p \\
&\simeq J_p(I) +\frac{1}{2} i^{T}\cdot Q\cdot i+S^T\cdot i
\end{align}

where $S = p(I_1^{p-1},\ldots, I_M^{p-1})$ and $Q =\operatorname{diag}\big(p(p-1)(I_1^{p-2},\ldots I_M^{p-2}) \big)$.  Thus, for a given initial value of $I$, problem (\ref{Eq:Opt}) becomes

\begin{equation}\label{Eq:Opt}
	\begin{aligned}
		& \underset{I}{\text{minimize}} & & \tilde{J}_p(i)  = \frac{1}{2} i^{T}\cdot Q\cdot i+S^T\cdot i\\
		& \text{subject to} & & C\cdot i = 0
	\end{aligned}
\end{equation}

We can write $\tilde{J}_p(i)$ as
\begin{align}
\tilde{J}_p(i) &= \sum_{m=1}^{M} = \frac{1}{2}p(p-1)I_{m}^{p-2}i_m^2+pI_m^{p-1}i_m \\
&= \sum \frac{1}{2}p(p-1) I_m^{p-2}\bigg( i_m^2+\frac{2I_m}{p-1}\bigg)\\
&=\sum\frac{1}{2}p(p-1)I_m^{p-2}\bigg(\big(i_m+\frac{I_m}{p-1}\big)^2 - \frac{I_m^2}{(p-1)^2} \bigg)
\end{align}
defining a new variable $\hat{i}_m = i_m+\frac{I_m}{p-1}$ we have the following optimization problem

\begin{equation}\label{Eq:Opt}
	\begin{aligned}
		& \underset{I}{\text{minimize}} & & \tilde{J}_p(\hat{i})  = \sum_{m=1}^{M} r_m \hat{i}_m^2\\
		& \text{subject to} & & C\cdot \hat{i} = \frac{1}{p-1}CT
	\end{aligned}
\end{equation}
where $r_m = \frac{1}{2}p(p-1)I_m^{p-2}$.
The problem in this canonical form is easy to solve distributedly using Jacobi iterations. For each node we define a potential $u$ such that if nodes $r$ and $s$ with $r<s$ are connected by edge $m$ then we have
\begin{align}
\hat{i}_m = \frac{u_r-u_s}{r_m}
\end{align}
Each node calculates its potential only as a function of the potentials of its adjacent nodes. Let $A_n$ be the set of nodes that are adjacent to node $n$. Let $\xi_n(k)$ be the index of the link that connects node $n$ to an adjacent node $k$. Then by Jacobi iterations we have

\begin{align}
u_n = \frac{\frac{[CT]_n}{p-1}+\sum_{k \in A_n} \frac{u_k}{r_{\xi_n(k)}}}{\sum_{k\in A_n}\frac{1}{r_{\xi_n(k)}}}
\end{align}. Note that the potential for the destination (node $N$) is always zero. After calculating the potentials for all of the nodes it is $i_m$ is

\begin{align}
i_m = \frac{u_r-u_s}{r_m} - \frac{I_m}{p-1}
\end{align}.

To summarize, up to here we calculated distributedly a correction vector $i$ so that changing the traffic loads from $I$ to $I + i$ reduces the cost function $J_p(I)$. Since $J_p$ is a convex function of $I$ for $p>1$, this SQP steps can be repeated several times in order to reach the global minimum of $J_p$. Note that as $p\to \infty$ the optimization problem becomes a minimax problem trying to distribute the traffic load evenly in the minimum cut-set links.

\end{document}